\documentclass[10pt]{sig-alternate-05-2015}
\CopyrightYear{2016}
\setcopyright{acmcopyright}
\conferenceinfo{ACM-ICN '16,}{September 26--28, 2016, Kyoto, Japan}
\isbn{978-1-4503-4467-8/16/09}
\acmPrice{\$15.00}
\doi{http://dx.doi.org/10.1145/2984356.2984376}

\begin{document}
\setcopyright{acmcopyright}
\title{Securing Content Sharing over ICN}

\numberofauthors{2}

\author{
\alignauthor Nikos Fotiou$^{\small{\mbox{*}}}$\\
        \affaddr{~}\\
        \affaddr{$^{\small{\mbox{*}}}$Mobile Multimedia Lab, Dept. of Informatics}\\
        \affaddr{\mbox{School of Information Sciences and Technology}}\\
        \affaddr{\mbox{Athens University of Economics and Business}}\\
        \affaddr{Evelpidon 47A, 113 62 Athens, Greece}\\
        \affaddr{~}\\
      \email{fotiou@aueb.gr}
\alignauthor George C. Polyzos$^{{\small{\mbox{*}}},{\scriptsize{\mbox{\#}}}}$\\
        \affaddr{~}\\
    \affaddr{$^{\scriptsize{\mbox{\#}}}$\mbox{Dept. of Computer Science and Engineering}}\\
    \affaddr{Jacobs School of Engineering}\\
        \affaddr{University of California, San Diego}\\
        \affaddr{La Jolla, CA 92093-0404, USA}\\
        \affaddr{~}\\
    \email{polyzos@acm.org}
}

\maketitle
\begin{abstract}
The emerging Information-Centric Networking (ICN) paradigm
is expected to facilitate content sharing among users.
ICN will make it easy for users to appoint storage nodes,
in various network locations, perhaps owned or controlled by them,
where shared content can be stored and disseminated from. 
These storage nodes should be (somewhat) trusted
since not only they have (some level of) access to user shared content,
but they should also properly enforce access control.
Traditional forms of encryption introduce significant overhead when it 
comes to sharing content with large and dynamic groups of users.
To this end, proxy re-encryption provides a convenient solution.
In this paper, we use Identity-Based Proxy Re-Encryption (IB-PRE)
to provide confidentiality and access control for content items shared over ICN,
realizing secure content distribution among dynamic sets of users.
In contrast to similar IB-PRE based solutions, our design allows each user 
to generate the system parameters and the secret keys required by the underlay encryption scheme
using their own \emph{Private Key Generator}, therefore,
our approach does not suffer from the key escrow problem.
Moreover, our design further relaxes the trust requirements on the storage nodes 
by preventing them from sharing usable content with unauthorized users.
Finally, our scheme does not require out-of-band secret key distribution.
\end{abstract}
\ccsdesc[500]{Networks~Network security}
\ccsdesc[500]{Networks~Security protocols}
\ccsdesc[500]{Networks~Middle boxes / network appliances}
\ccsdesc[300]{Networks~Network architectures}
\ccsdesc[300]{Networks~Network protocols}

\ccsdesc[500]{Security and privacy~Security protocols}
\ccsdesc[500]{Security and privacy~Authentication}
\ccsdesc[500]{Security and privacy~Access control}
\ccsdesc[500]{Security and privacy~Authorization}
\ccsdesc[300]{Security and privacy~Cryptography}
\ccsdesc[300]{Security and privacy~Security services}
\ccsdesc[100]{Security and privacy~Network security}

\printccsdesc
\keywords{
Identity-based encryption, Proxy re-encryption,
Content-Centric Networking,
Information-Centric Networking,
Named-Data Networking
}
\section{Introduction}
Internet users regularly utilize web-based systems to share content with others.
These systems provide facilities that allow content owners to store shared items online,
as well as, to define access control policies. Authorized third parties can access the 
shared content, even if the content owner is disconnected.
The emerging Information-Centric Networking (ICN) paradigm
can be an excellent technology for implementing a content sharing system.
Being information oriented, ICN, can achieve significant gains in terms
of network performance and offers opportunities for innovative security solutions.
%

An important aspect of content sharing systems is that they should be highly trusted
as they have access to user content and they are responsible for enforcing access control policies.
Encrypting content before uploading it to a storage service can assuage 
most of the security and privacy concerns.
Nevertheless, traditional forms of encryption are not convenient for sharing content with large and dynamic groups
of users, as they require complex key management systems and pose significant 
communication and/or storage overhead when a user is added or removed from the list of authorized users.
However, the same goal can be achieved more efficiently by using \emph{Proxy Re-Encryption} (PRE)~\cite{Gre2007}.
PRE allows an entity, to transform a ciphertext computed under 
Alice's public key into one that can be decrypted by Bob's secret key,
without having access neither to Alice's or Bob's secret keys, nor to the un-encrypted content.
Therefore, by using PRE, a content owner can encrypt each content item only once
and provide the storage node with the proper re-encryption keys, as well as,
with the desired access control polices.
In this setup the storage node needs to only be trusted to re-encrypt content only for authorized users (as determined by the access control policies) for proper system operation.
But even if it fails to properly perform the re-encryption, no usable content is revealed.

In this paper, we use Identity-Based Proxy Re-Encryption (IB-PRE) to implement a security
and access control system for  ICN-based content sharing and
introduce the following
two constructions.

Our first construction is a traditional IB-PRE-based system, with the added advantage that 
it allows users to use their own \emph{Private Key Generators} (PKG), i.e., each user is able 
to generate by himself his secret key.
This is a significant advantage compared to existing systems where a 
centralized PKG generates the keys for all users and therefore these 
systems suffer from the key escrow problem, i.e., PKGs know the private keys of the users and therefore they can impersonate them.

Our second construction improves on the first by relaxing the requirement for semi-trusted proxies,
in the sense that our scheme is secure even if the storage node does not respect access control policies.
With this construction a storage node can re-encrypt a content item \emph{only} for authorized users.

In addition to these constructions, we provide a framework for defining and managing simple 
access control policies, as well as, an authentication protocol that can be used for user-to-storage node 
authentication as well as for \emph{content-to-user authentication}, i.e.,
it allows end-users to verify that they received the desired content item from an \emph{authorized}
storage node. Our schemes do not require out-of-band secret key distribution and have very low overhead.  

The remainder of
this paper is organized as follows.
In Section~\ref{sec:back} we review identity-based encryption and
proxy re-encryption and we discuss related work in the area.
In Section~\ref{sec:syst} we detail our system design.
In Section~\ref{sec:eval} we present our implementation and we evaluate our solution.
In Section~\ref{sec:discuss} we discuss various design choices and alternatives and the related tradeoffs.
Finally, in Section~\ref{sec:conc} we 
conclude with a summary of the contributions of this paper.

\section{
Background \& Related Work}
\label{sec:back}
\subsection{Identity-Based Encryption}
An Identity-Based Encryption (IBE) scheme is a public key encryption 
scheme in which an arbitrary string can be used as a public key.
An IBE scheme is specified by four algorithms:
\texttt{Setup}, \texttt{Extract}, \texttt{Encrypt} and
\texttt{Decrypt}.
\begin{itemize}
\item \texttt{Setup}: it is executed by a Private Key Generator (PKG).
It takes as input a security parameter $k$
and returns a \texttt{master-secret key} ($MSK$) and some \texttt{system parameters} ($SP$).
The $MSK$ is kept secret by the PKG, whereas $SP$ are made publicly available.
\item \texttt{Extract}: it is executed by a PKG. It takes as input $SP$, $MSK$, and an arbitrary string $ID$,
and returns a \texttt{secret key} $SK_{ID}$. 
\item \texttt{Encrypt}: takes as input an arbitrary string $ID$, a message $M$, and $SP$,
and returns a ciphertext $C_{ID}$.
\item \texttt{Decrypt}: takes as input $C_{ID}$, the corresponding private decryption key $SK_{ID}$,
and returns $M$.
\end{itemize}
Therefore, using IBE and providing that $SP$ are known,
it is possible to encrypt some plaintext using an arbitrary $ID$ as the public key.
The entity that holds the $SK$ that corresponds to this $ID$ can decrypt the ciphertext.

\subsection{Proxy Re-Encryption}
A Proxy Re-Encryption (PRE) scheme is a scheme in which a third semi-trusted party,
called, is allowed to alter a ciphertext, encrypted with the public key of a user $A$ (the delegator),
in a way that another user B (the delegatee) can decrypt it with her own appropriate key
(i.e., in most cases her secret private key).
Green and Ateniese~\cite{Gre2007} implemented an Identity-Based Proxy Re-Encryption (IB-PRE) scheme
that specifies two new algorithms, \texttt{RKGen} and \texttt{Reencrypt},
in addition to the IBE algorithms already discussed.
\begin{itemize}
\item \texttt{RKGen}: it is executed by the owner of a public key $ID1$.
It takes as input $SP$, his secret key $SK_{ID1}$, and an identity $ID2$
and generates a (public) re-encryption key $RK_{ID1\rightarrow ID2}$.
\item \texttt{Reencrypt}: it is executed by a third semi-trusted party.
It takes as input $SP$, a re-encryption key $RK_{ID1 \rightarrow ID2}$, and a ciphertext $C_{ID1}$
and outputs a new ciphertext $C_{ID2}$.
\end{itemize}
\begin{figure}
\includegraphics[width=1.0\linewidth]{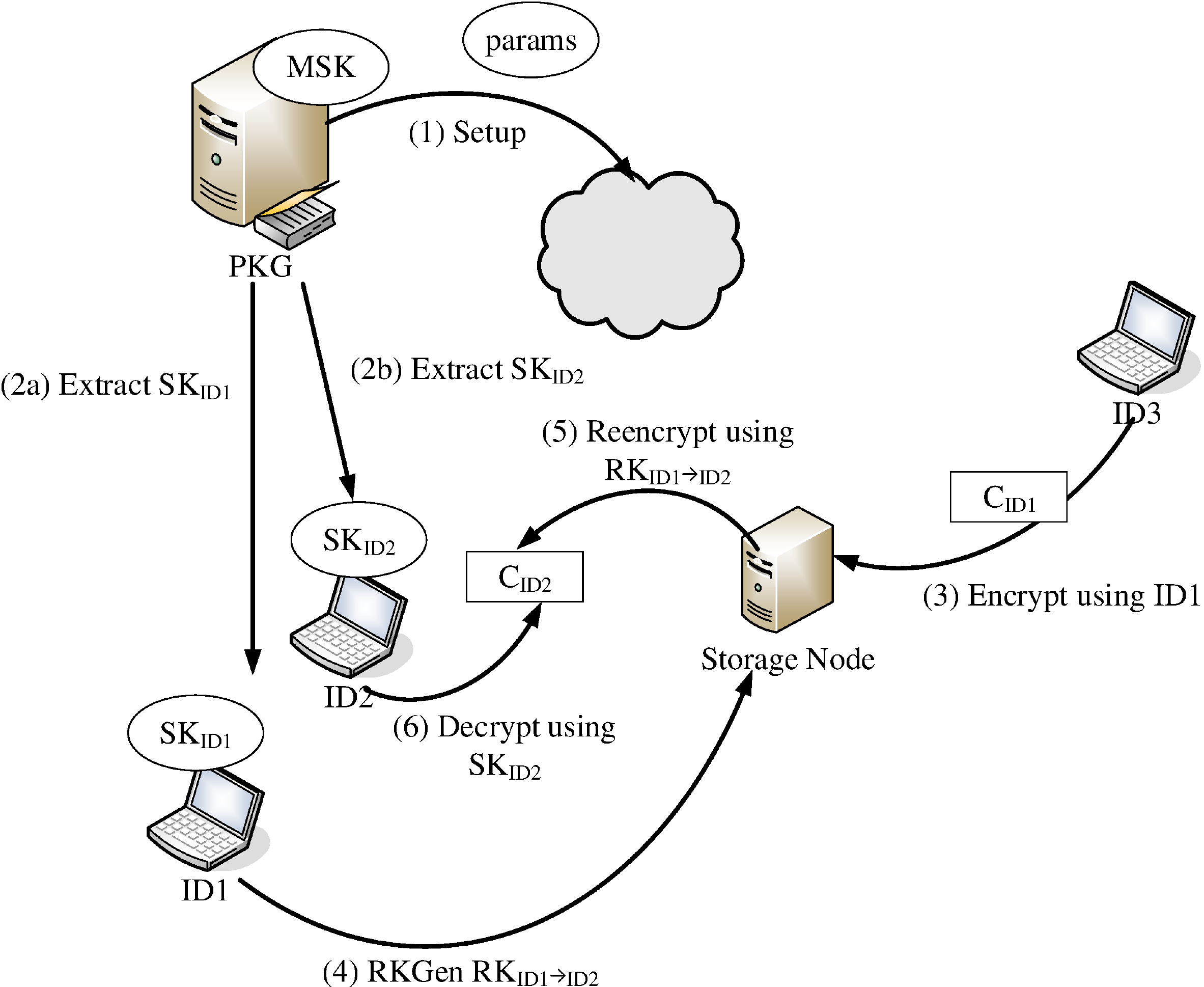}
\caption {IBE-PRE example}
\label{fig:ibe-pre}
\end{figure}
Figure~\ref{fig:ibe-pre} gives an example of a complete IBE-PRE transaction.
In this figure, initially the PKG generates the $MSK$ and the $SP$ and makes
the $SP$ publicly available (step 1).
Then it extracts $SK_{ID1}$ and $SK_{ID2}$ and distributes them to the corresponding users (step 2).
Another user creates a ciphertext using as a public key the string $ID1$ and stores it in a storage node (step 3).
The user that owns $ID1$ creates a re-encryption key $RK_{ID1 \rightarrow ID2}$ and sends it to the storage node.
The node, re-encrypts $C_{ID1}$ using $RK_{ID1 \rightarrow ID2}$ and generates $C_{ID2}$.
The user that owns $ID2$ is now able to decrypt the new ciphertext.
The storage node learns nothing about the contents of the ciphertext or the secret keys of the users.

\subsection{Related work}
Early attempts to secure distributed content sharing systems  suffered from key management problems.
In these systems (e.g., Plutus~\cite{Kal2003}) each content item is encrypted with a symmetric encryption key.
Then this key has to be (securely) delivered individually, to each authorized user. 
Therefore, a large number of keys has to be stored and transferred to storage nodes.
Our system exhibits much lower communication and storage overhead and uses simpler key management.    

Advances in cryptography allowed the development of more efficient systems.
A cryptographic scheme that has received significant attention by the research community is \emph{Attribute Based Encryption} (ABE).
ABE allows the specification of ``attributes'' that a user should have in order to decrypt a ciphertext.
This is achieved by associating attributes with private keys.
In addition, ABE allows the definition of simple access control policies using logical operators,
e.g. a ciphertext can be decrypted  by the users who have the `CS' AND `student' OR `professor' attributes.
ABE has been widely used for implementing contemporary systems
with distributed content sharing capabilities (e.g.,~\cite{Bad2009},\cite{Nil2012})
and has received considerable attention by the ICN community (e.g.,~\cite{Ion2013},\cite{Li2014},\cite{Sil2015}) .
In these systems users use attributes to describe \emph{groups} of trusted peers 
(e.g., `friends', `colleagues' etc.), then they encrypt their content items using a symmetric encryption key
and encrypt this key using ABE.

The main drawback of these ABE systems is that removing a group member requires re-keying,
i.e., the private keys that correspond to each attribute have to be regenerated and re-distributed. 
The same problem exists when a (compromised) private key has to be revoked.
Jahid et al.~\cite{Jah2011} mitigate this problem by using the key revocation scheme presented in~\cite{Mon2001}.
With this scheme the decryption of a ciphertext requires two keys:
the private key that corresponds to the user's attributes
and a secondary that is provided by a semi-trusted third party.
The third party generates the secondary key for non revoked users using a secret key.
Every time a user is removed from a group or a key has to be revoked,
the only key that has to be updated is the secret key.
In our work
access control and user management are decoupled from the encryption of the shared content item,
therefore, adding or removing a user from an access control policy
does not involve any new encryption of the protected items.
In addition, in ABE-based schemes
content owners should securely distribute to other users the private keys that correspond to each attribute.
In our design no out-of-band secret key distribution is required.

Zhang et al.~\cite{Zha2011} utilize the IBE scheme proposed by Boneh and Franklin~\cite{Bon2001}
and the identity based signature scheme proposed by Hess~\cite{Hes2003}
in order to provide name-based trust for the NDN architecture~\cite{Jac2009}.
The identity used  for the encryption of a content item in this scheme is either the name of the shared item,
or the identity of the item's recipient.
In the former case, the item's recipient learns the $SK$ that corresponds to the content name using out-of-band mechanisms.
This solution is focused on content confidentiality and it is not efficient when it comes to access control.
Our system combines proxy re-encryption with IBE
and provides efficient access control in addition to content confidentiality.

Ateniese et al.~\cite{Ate2006} use IB-PRE for implementing a secure storage service.
In this solution content items are stored in a semi-trusted storage node,
encrypted with a symmetric key.
The symmetric key is encrypted with the identity of the content owner. 
Every time a user requests access to a content item,
the storage node requests from an access control server to create a re-encryption key,
which is used for re-encrypting the symmetric key in a way that the content requester can decrypt it.
The proposed solution uses a single $PKG$.
This solution is used in an ICN context by Wood and Uzun~\cite{Woo2014}
to implement a DRM-like solution for the CCN architecture as well as
by Zheng et al.~\cite{Zhe2015} to implement an access control mechanism for ICN.
Our work extends these previous works by considering user-specific $PKGs$.
Moreover, in these prior solutions, once the storage node learns the re-encryption key for a user,
it may use it to re-encrypt all other shared items of the content owner,
no matter whether individual users are authorized to access them or not.
Our solution includes a construction that mitigates this problem.

\section{System design}
\label{sec:syst}
\subsection{Inter-domain IB-PRE}
IB-PRE solutions usually assume that delegators and delegatees
belong to the same \emph{administrative domain},
therefore, they share the same PKG.
This setup in many cases is not realistic.
Moreover,
given that PKGs know the users' secret keys,
grouping many users under the same PKG raises security concerns.
All these problems can be mitigated using \emph{Inter-domain IB-PRE}~\cite{Tan2009}.
Inter-domain IB-PRE allows a $3^{rd} party$ to transform
a ciphertext computed using an $ID$ and the $SP$ of a domain $A$ into
a ciphertext intended for an $ID$ and the $SP$ belonging to another domain $B$.
The Green-Ateniese scheme implements Inter-domain IB-PRE by allowing the usage of the
$SP$ of the delegatee's domain as input to the \emph{RKGen} algorithm.\footnote{
As a matter of fact, the Green-Ateniese scheme can even transform an IBE ciphertext into an 
RSA ciphertext (see section 5 of~\cite{Gre2007}). }
  
\subsection{Setup}
\label{setup}
Our system considers two types of users: content \emph{owners} and content \emph{subscribers}.
All users are uniquely identified by an \emph{identity}.
The form and the semantics of this identity are user specific
(e.g., an email address, a domain name, or a real world name).
Moreover, diverse identity types may co-exist in the same instance of our system.
Each user maintains his own PKG  which is used to generate the user specific $SP$,
as well as, the $SK$ that corresponds to his identity.
A resolution service that maps identities to $SP$s is assumed.
Such a service could be implemented using DANE~\cite{Hof2012}, Blockchains~\cite{Fot2016}, Keyservers,
third party directories, or even out-of-band mechanisms.
Owners appoint (or own) \emph{storage nodes},
where lists of known subscribers, access control policies, and shared content are stored.
Storage nodes act as ICN publishers and are responsible for enforcing access control policies
and for forwarding (encrypted) content items to \emph{authorized} subscribers.
Shared content items are identified by ICN routable identifiers,
which are ``advertised''  in the network by the storage nodes
using standard ICN procedures.

It should be noted that our system does not consider publisher (i.e. storage node) identities,
since subscribers are interested in receiving a piece of content no matter its location.
In section~\ref{hand} we present a protocol that allows
a storage node to prove to a subscriber
that \emph{it is authorized to store a particular content item}.
 
\subsection{A first construction}
Each content owner maintains at a storage node three data structures:
a table of \emph{Known Subscribers}, a table of \emph{Access Control Policies}, and a table of \emph{Shared Content}. 
The table of \emph{Known Subscribers} contains rows of the form $[Identity, SP, RK]$,
where $Identity$ is the identity of a subscriber known to the content owner, $SP$ is the $SP$ of that subscriber,
and  $RK$ is the re-encryption key $RK_{onwer \rightarrow identity}$,
where \emph{owner} is the identity of the content owner
and \emph{identity} is the identity located in the first column of the row.
The table of \emph{Access Control Policies} contains rows of the form $[Policy, List{<}Identity{>}]$,
where $Policy$ is an identifier of the access control policy
and $List{<}Identity{>}$ is a list of subscriber identities that abide by that access control policy.
These identities should also exist in the \emph{Known Subscribers} table.
Finally, the table of \emph{Shared Content} contains rows of the form $[Item, Policy, C_{owner}(K)]$,
where $Item$ is the identifier of the (shared) content item,
$Policy$ is the policy of the access control policy that protects this item,
and $C_{owner}(K)$ is the encryption of the symmetric encryption key $K$,
generated as described in the following section.

For each content item she wants to share, a content owner, selects a symmetric encryption key $K$,
encrypts the item using a symmetric encryption algorithm $Enc$,
and encrypts $K$ using the \emph{Encrypt} IBE algorithm with input her $SP$ and her identity,
producing this way $C_{owner}(K)$.\footnote{In the Green-Ateniese scheme the message $M$ that is used
in the \emph{Encrypt} algorithm is an element of a group $G$ of prime order $q$.
Therefore, to be precise, the content owner selects a random element $K$ in $G$ and uses a secure hash function $H:G \rightarrow \{0,1\}^n$ (where $n$ is sufficiently large, e.g., 128) to calculate $H(K)$,
which is then used as a key by the symmetric encryption algorithm.}
Then she stores the content item in a storage node and populates its tables accordingly.

\begin{figure}
\begin{center}
\includegraphics[width=0.95\linewidth]{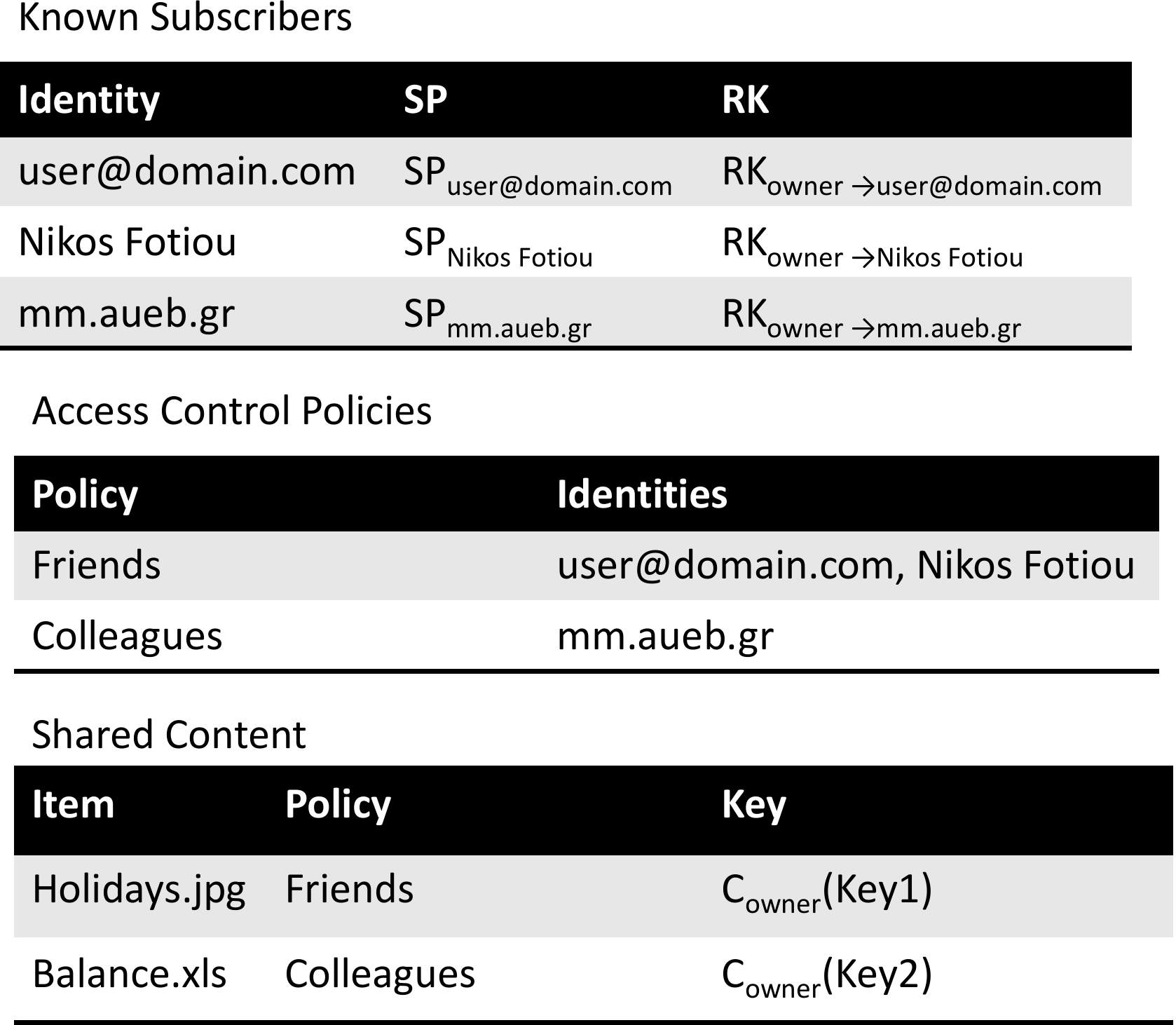}
\caption {Storage node components with the first construction }
\label{fig:proxy}
\end{center}
\end{figure}

Figure~\ref{fig:proxy} illustrates an instance of a storage node.
As it can be seen, the \emph{Known Subscribers} table of the node contains three
subscriber identities.
For each identity the corresponding re-encryption key has been created.
Moreover, two access control policies have been created using the subscriber identities.
Finally, two items have been added in the table of \emph{Shared Content}.

\subsubsection{Content sharing with trusted subscribers}
A content item can be shared with some of the subscribers included in the table of \emph{Known Subscribers}.
A content owner should perform the following steps:
(i)~create an access control policy,
assign to this policy a list of subscriber identities that can access the item,
and update accordingly the storage node's \emph{Access Control Policies} table,
(ii)~encrypt the item using a symmetric encryption key and encrypt this key using her identity,
i.e., generate $C_{owner}(K)$, and 
(iii)~copy the encrypted item in the storage node and update its \emph{Shared Content} table.

\subsubsection{Content sharing with subscribers another known subscriber knows}
In many cases it is desirable to share a content item with subscribers another known subscriber knows, e.g., 
in terms of social networks with the `friends' of my friend.
In order to implement this functionality, well-known policy identifiers can be used.
Suppose, an owner with identity $A$ trusts a subscriber with identity $B$,
and $A$ wants to share a content item with the `colleagues' of $B$.
In this case `colleagues' is a well-known policy identifier.
$A$ encrypts the item using symmetric encryption and encrypts the symmetric encryption key
using IBE \emph{encrypt} with input $B$ and \emph{$B's$ $SP$}.\footnote{Remember
that $SP$ are publicly available.}
I.e., $A$  creates $C_{B}(K)$ using  $B's$ $SP$.
Then, the encrypted item, as well as $C_{B}(K)$, are stored \emph{in $B's$ storage node},
and $B's$ storage node table of \emph{Shared Content} is modified accordingly.
It should be noted that since the symmetric encryption key is encrypted using $B$ as the public key and $B's$ $SP$, $B$ can also access the content item.

\subsubsection{Endpoints authentication and secure channel setup}
\label{hand}
Our construction assures that subscribers that ``lie'' about their identity
cannot decrypt the received content.
Therefore, having a storage node blindly accepting a subscriber's claims
about his identity can be a valid design choice.
On the other hand it may be 
desirable to have communicating endpoints authenticated to each other.
E.g., for preserving bandwidth by not sending content items to unauthorized subscribers, 
for hiding the existence of a content item from unauthorized subscribers, for hiding subscribers' 
interests from malicious storage nodes, etc.,
as well as to transmit control messages over a secure communication channel.

In the following we present a handshake protocol that takes place before any content request and has the following properties (i) it provides subscriber authentication, (ii) it provides a proof that a storage node is authorized to store a particular content item, and (iii) it enables the creation of an ephemeral symmetric encryption key that can be used for securing the communication channel between the two endpoints. 
 
Let $U$ be the identity of a subscriber that wants to access an item $F$.
Moreover, let $H$ be an 
keyed-Hash Message Authentication Code (HMAC)
function, and $Enc$ a symmetric encryption algorithm.\\
%
\textbf{step 0:} The content owner uses the \emph{Extract} IBE algorithm, generates $SK_{F}$, i.e, 
the secret key that corresponds to the item identifier, and stores this key to the storage node that hosts the item.\\
\vspace{-0.5\baselineskip}
\\
\textbf{step 1:} $U$ learns out-of-band the owner's $SP$ and generates a random 
number $r$, a key $h$ that is used by $H$, a key $k$ that is used by $Enc$, and a Diffie-Hellman parameter $DH_{U}$, 
and sends to the storage node the following subscription message:
$$msg1: F, C_{F}(k), Enc_k(r,h,DH_{U},U), H_h(MSG)$$
where $C_F$ is a ciphertext generated using the IBE \emph{encrypt} algorithm 
with input $F$ and $k$ and $H_h(MSG)$ is the output of $H$ using $h$ applied over the whole message.
The content item identifier $F$ is included in the plaintext for two reasons:
first, because it is required  by the underlay ICN routing plane and, second,
to help shared storage nodes to decide which $SK$ to use.
(Note: this is similar to the Server Name Indication TLS extension~\cite{Eas2011}.)\\
\vspace{-0.5\baselineskip}
\\ 
\textbf{step 2:} The storage node ($Node$) decrypts $C_{F}(k)$ (using $SK_{F}$ from step 0),
 then decrypts $Enc_k(r,h,DH_{U},U)$, and then verifies $H_h(MSG)$.
If the verification succeeds, the storage node retrieves $U$'s $SP$ from the  \emph{Known Subscribers} table, 
generates a random number $r'$, a key $k'$, and a Diffie-Hellman parameter 
$DH_{Node}$, and sends to $U$:
$$msg2: C_U(k'), Enc_{k'}(r,r',DH_{Node}), H_h(MSG)$$
\textbf{step 3:} $U$ verifies $H_h(MSG)$; if the verification succeeds, $U$ decrypts $C_U(k')$, then decrypts $Enc_{k'}(r,r',DH_{Node})$, and then verifies that $r$ is the same with that included in his first message.
If the verification succeeds, $U$ calculates a secret key $s$ using $DH_{U}$ and $DH_{Node}$, and sends to the storage node: 
$$msg3: Enc_{s}(r'), H_h(MSG)$$
\textbf{step 4:} The storage node verifies $H_h(MSG)$; if the verification succeeds, it calculates $s$, it decrypts
$Enc_{s}(r')$, and examines if $r'$ is the same with that included in its previous message.

At the end of this protocol, the subscriber has been authenticated to the storage node,
the storage node has proven to the subscriber that it is authorized to host the desired content item, and
both entities have established common encryption and HMAC keys.\\
\vspace{-0.5\baselineskip}
\\ 
\textbf{Proof:} $C_{F}$ can only be decrypted by an entity that knows $SK_{F}$.
If $msg1$ and $msg2$ contain the same values for $r$ then the storage node knows $SK_{F}$.
Similarly $C_U$ can only be decrypted by the owner of $SK_U$.
If $msg2$ and $msg3$ contain the same value for $r'$ then the subscriber knows $SK_U$.
The hashes included in all messages protect the messages integrity.
Moreover, providing that $U$ learns the correct storage node's $SP$, man in the middle attacks are not possible.

As a next step,
$U$ issues a  new subscription message that includes in its payload a \emph{content request}.
The request is encrypted with $ENC_{s}$ and the hash of the ciphertext is calculated using $H_h$.  
Upon receiving the request, the storage node
(i) retrieves the record that corresponds to the content identifier included 
in the request from the \emph{Shared Content} table,
(ii) retrieves the list of identities that abide by the access control policy
used to protect the item in question from the \emph{Access Control Policies} table, 
(iii) checks if $U$ is included in the list of identities retrieved during step (ii); if yes,
(iv) it retrieves the re-encryption key that corresponds to $U$ from the 
\emph{Known Subscribers} table, re-encrypts $C_{owner}(K)$ (i.e., the encrypted
key found in the \emph{Shared Content} table), and sends the new 
ciphertext along with the encrypted content item to $U$.

\subsection{A second construction: allowing untrusted storage nodes}
In the first construction, and generally in most proxy re-encryption schemes,
the storage node is trusted to perform the re-encryption process only for authorized subscribers.
A malicious storage node, however, can use a re-encryption key to re-encrypt all $C_{owner}(K)$ ciphertexts,
and make them accessible to a subscriber no matter whether the subscriber is authorized to access $K$.
This attack can be mitigated using the following construction.

For each access control policy identifier,
the content owner uses the \emph{Extract} IBE algorithm to generate $SK_{Policy}$.
Then, for each content item she wants to share, she selects a symmetric encryption key $K$,
encrypts the item using a symmetric encryption algorithm and $K$,
and encrypts $K$ using the \emph{Encrypt} IBE algorithm
with input her $SP$ and the \emph{identifier of the access control policy that protects the content item},
producing this way $C_{Policy}(K)$.

\begin{figure}
\begin{center}
\includegraphics[width=0.95\linewidth]{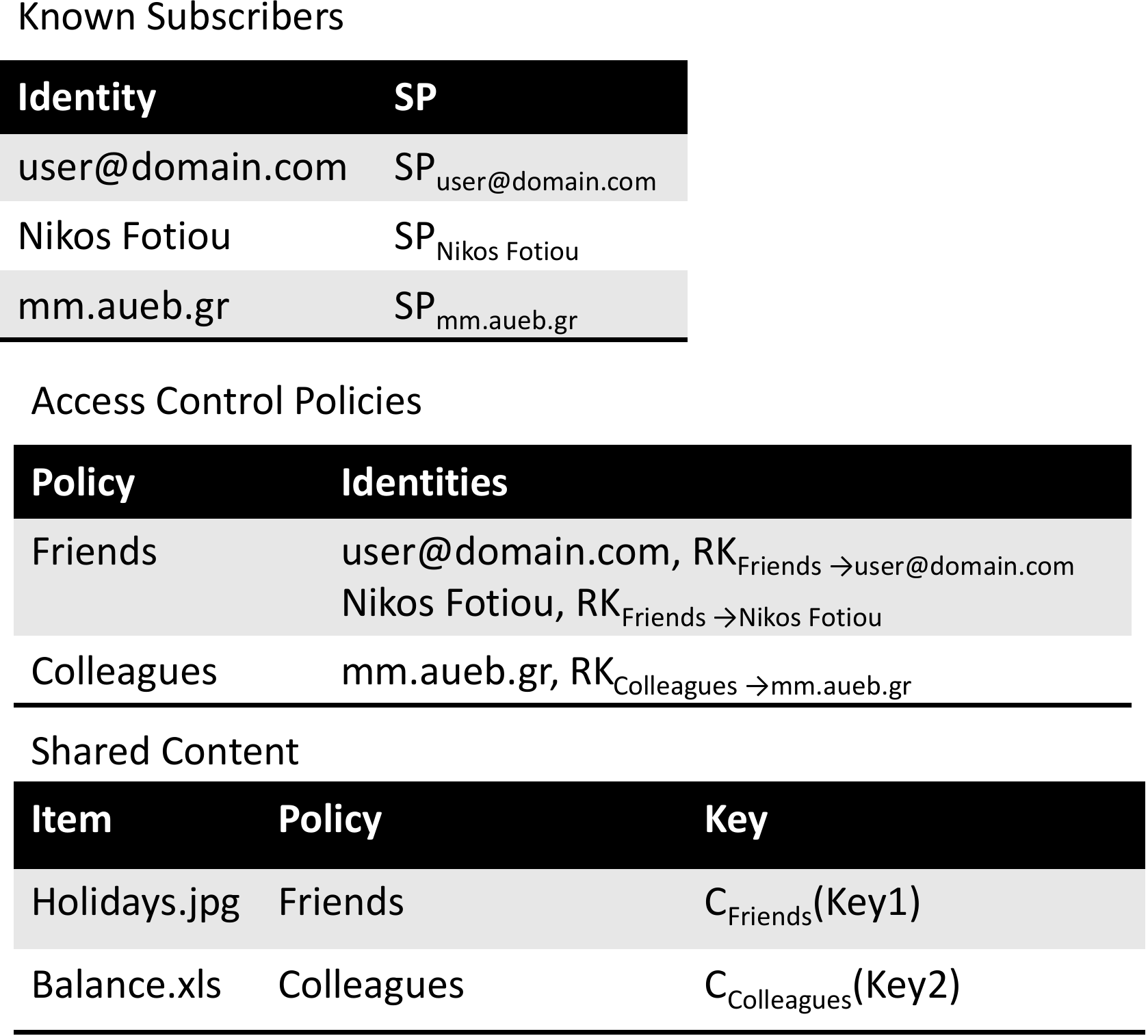}
\caption {Storage node components with the second construction }
\label{fig:proxy2}
\end{center}
\end{figure}

The data structures of the storage nodes are now modified as follows.
The table of \emph{Known Subscribers} contains rows of the form $[Identity, SP]$.
The table of \emph{Access Control Policies} contains rows of the form $[Policy, List{<}(Identity, RK){>}]$,
where $List{<}(Identity, RK){>}$ is a list of pairs of subscriber identities that abide by that access control policy
and the corresponding re-encryption key $RK_{Policy \rightarrow identiy}$ generated using the $SK_{Policy}$.
Finally, 
the table of \emph{Shared Content} contains rows of the form $[Item, Policy, C_{Policy}(K)]$.
Figure~\ref{fig:proxy2} illustrates an instance of a storage node based on the second construction.

When a subscriber $U$ subscribes to a content item,
the storage node selects $RK_{Policy \rightarrow U}$ from the \emph{Access Control Policies} table
to re-encrypt $C_{Policy}(K)$.
This re-encryption key can only be used to re-encrypt ciphertexts encrypted using $Policy$ as the public key, i.e., ciphertexts that $U$ is authorized to access. 

It should be noted that these constructions are compatible with each other,
i.e., an owner that uses the first construction can share a content item with a subscriber
that uses the second construction and vice versa.

\section{Evaluation}
\label{sec:eval}
\subsection{Security evaluation}
The Green-Ateniese scheme has been proven to have the following security properties~\cite{Gre2007}:
(i)~it is chosen-plaintext attack (CPA) secure,
(ii)~re-encryption key generation does not require any delegatee involvement,
(iii)~the entity that performs a re-encryption can not create new re-encryption keys from the existing ones, and
(iv)~re-encryption keys are \emph{unidirectional}.

Each content item is encrypted using a different symmetric encryption key,
therefore, the compromise of a symmetric encryption key
would require a new encryption of the content item with another fresh key
and the update of the corresponding entry in the \emph{Shared Content} table.
This is an inevitable overhead of all similar systems and it is due to the fact that public key encryption cannot be applied directly to the content item,
because of its computational complexity.
Nevertheless, if small items are considered,
e.g., sensor measurements, news headlines, tweets, etc.,
it could be possible to relax the need for symmetric encryption.

Access control policy management in our system is orthogonal to content encryption.
Let $A$, $B$, $C$ be three users,
and suppose that $A$ has created an access control policy named `friends' that includes $B$ and $C$.
If $A$ decides to remove $B$ from this policy,
he has simply to update the corresponding entry in the \emph{Access Control Policies} table.
No further action is required (e.g., no re-keying or re-encryption of a content item), provided
that the storage node will not use any cached re-encryption key generated
while $B$ was authorized.
This is not the case with systems that incorporate access control policies in ciphertexts
(e.g, ABE-based systems);
in these systems the modification of an access control policy requires the
re-encryption of the content item (or of the symmetric key that protects this item).

If the master secret key of user $A$'s PKG is compromised,
$A$ has to perform the following actions:
(i)~generate new $SP$, master secret key, and $SK$s,
(ii)~create fresh re-encryption keys,
(iii)~create fresh $C_{A}(K)$ (or $C_{Policy}(K)$) ciphertexts, and
(iv)~update the corresponding storage node tables.
Content items are not required to be encrypted again.
Moreover, any other user that shares content with $A$ should update the corresponding re-encryption keys.

If the $SK$ of a user is compromised,
the user has two options:
either perform the same steps as in the case of master secret key loss,
or change identity (e.g., using the serial number approach of~\cite{Fot2015}).
With the latter option the user has only to update the re-encryption keys.
Finally, when it comes to our second construction,
in case of a $SK_{Policy}$ compromise,
a user has again the same two options:
either follow the procedure used for the master secret key loss,
or change the name of the access control policy
(the latter however may be harder for well known policy names).
Again, if a policy is renamed, the corresponding re-encryption keys should be updated.

A storage node learns nothing about subscriber secret keys or the shared content.
We now discuss the case of a storage node compromise.
We distinguish two types of attackers:
(a)~third party attackers, and
(b)~attackers that abide by a certain access control policy (i.e., authorized by the access control policy).
In the former case (i.e., attackers with no authorization at all),
the attacker gains no information about the stored content, neither about the subscriber secret keys. 
In the latter case  (i.e., attackers with authorization),
if our first construction is used,
the attacker is able to decrypt all content items by using the $RK_{onwer \rightarrow identity}$
re-encryption key included in the \emph{Known Subscribers} table;
if our second construction is used,
the attacker is able to decrypt only the items protected by the
so authorizing
access control policy.
In both cases, the attacker learns no information about the the secret keys of other users.

\subsection{Performance evaluation}
The IB-PRE algorithm of Green and Ateniese used in our system has been 
implemented\footnote{We used the first construction presented in~\cite{Gre2007}.}
using the Charm Crypto library~\cite{Aki2012}.
In order to achieve a security level equivalent to RSA with key size 1024 bits, 
the size of $SP$ is 2048 bits, the size of $C_{ID}(key)$ is 2048 bits,
and the size of a re-encryption key is 3072 bits. 
In an Ubuntu 12.04 desktop machine
running in a single core of an Intel i5-4440 3.1 GHz processor and with 2GB of RAM, 
the creation of $C_{ID}(key)$, where $key$ is a 128 bits symmetric encryption key, required 40~ms;
the creation of a re-encryption key required 20~ms,
the re-encryption of a ciphertext required 31~ms, and the decryption of an IBE ciphertext required 28~ms.
These numbers are
means
of 20 runs of the same experiment;
variation
among
experiments
was
negligible.  
It should be noted that IBE is used for encrypting, decrypting, and re-encrypting a symmetric encryption key,
therefore these measurements are independent of the size of the content item. 

We now compare the storage and communication overhead of our solution against 
a public key based \emph{trivial solution} and an \emph{ABE based solution}. We
do not consider the overhead of the handshake protocol described in section~\ref{hand}.
In the case of the trivial solution, all users are equipped with a self-generated public/private key pair.
Public keys are distributed using out-of-band mechanisms.
Each content item is encrypted with a symmetric encryption key
and this key is encrypted with the public keys of all subscribers that are allowed to access the item.
In the ABE-based solution, we consider each policy as an attribute. 
Every content onwer generates public/private keys that correspond to attributes and distributes them accordingly.
Each content item is encrypted with a symmetric key
and this key is encrypted using a key that corresponds to a specific attribute;
only subscribers that own the corresponding attribute keys can decrypt the symmetric key.
Hence, we are considering CP-ABE without ``policy trees.''

For our evaluation we consider the following scenario.
A content owner knows $U$ subscribers and has grouped them in $G$ access control policies.
Each policy contains $U_G$ subscribers.
(A subscriber may belong to multiple policies.)
Moreover, the owner shares $F$ content items, each of which is protected by a single access control policy.  

\subsubsection{Storage overhead}
\begin{table}
\centering
\begin{tabular}{ | l | p{4.25cm} | p{0.75cm} |}
\hline
{} & {} & Value\\
Symbol & Meaning & (bits)\\ \hline\hline
$|ID|$ &Size of a subscriber identity or of a content item identifier & 256\\ \hline
$|SP|$ &Size of system parameters & 2048\\ \hline
$|SK{a\rightarrow b}|$ &Size of a re-encryption key & 3072\\ \hline 
$|C_{identity}(K)|$ &Size of a symmetric encryption key encrypted using IBE & 2048 \\ \hline 
$|PK|$ &Size of public key & 1024 \\ \hline 
$|Enc(K)|$ &Size of an encrypted symmetric encryption key & 1024 \\ \hline 
$|ABE_{attr}(K)|$ & Size of a symmetric encryption key encrypted using 
ABE (we consider the construction of~\cite{Bet2007}) & 4096\\ \hline 
\end{tabular}
\caption{Notation and parameters for storage overhead evaluation}
\label{notation}
\end{table}
\begin{figure}
\includegraphics[width=1.0\linewidth]{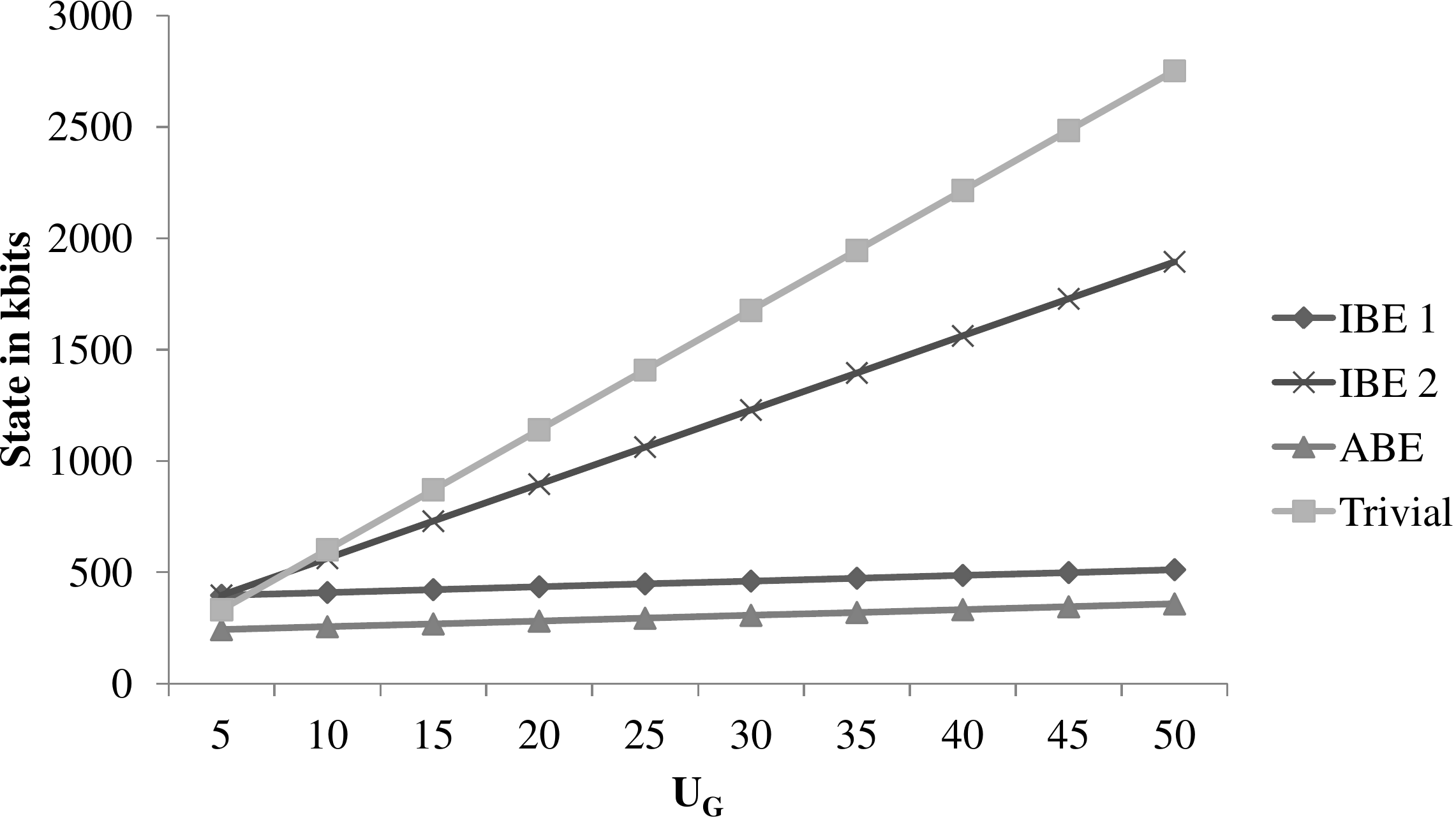}
\caption {Storage overhead as a function of $U_G$, the number of subscribers per access control policy}
\label{fig:state1}
\end{figure}

With our scheme, in addition to the encrypted content, a storage node maintains the following structures: 
a table of \emph{Known Subscribers}, a table of \emph{Access Control Policies}, and a table of \emph{Shared Content}.
In our evaluation scenario, and for our first construction, the storage node should maintain
for
$U$ identities $SP$ and re-encryption keys
(i.e., the \emph{Known Subscribers} table),
$G \times U_G$ identities in the Access Control Policies table, 
and $F$ content item identifiers and encrypted symmetric encryption keys.
For simplicity, we assume that content item identifiers and subscriber identities have the same size in bits. 
The storage overhead of our fist construction can then be calculated as follows:
\begin{align}
\begin{split}
S = {}& U \times (|ID| + |SP|+ |SK{a\rightarrow b}|)\\
      & + G \times U_g \times |ID|\\
      & + F \times (|ID|+|C_{owner}(K)|)
\end{split}
\end{align}
where $|ID|$ is the size of a subscriber identity (or of a content item identifier),
$|SK{a\rightarrow b}|$ the size of a re-encryption key,
and $|C_{owner}(key)|$ the size of the encrypted symmetric encryption key. 
Table~\ref{notation} contains this notation as well as the size of each field.

In our second construction re-encryption keys are stored in the 
\emph{Access Control Policies} table
instead of the \emph{Known Subscribers} table.
Therefore, the storage overhead can be calculated as follows:
\begin{align}
\begin{split}
S = {}& U \times (|ID| + |SP|) \\ 
  & + G \times U_g \times (|ID| + |SK{a\rightarrow b}|) \\
  & + F \times (|ID|+|C_{ACP}(K)|) 
\end{split}
\end{align}
When the trivial solution is used, then a storage node should maintain $U$ identities and public keys, 
$G \times U_G$ identities for the access control policies,  $F$ content item identifiers,
and $F \times U_G$ encryptions of the symmetric encryption key.
Therefore, in this case the size of the state maintained can be calculated as follows:
\begin{align}
\begin{split}
S ={} & U \times (|ID| + |PK|)\\ 
& + G \times U_G \times |ID| \\
& +F \times (|ID| +U_G \times |Enc(K)|)
\end{split}
\end{align}
where $|PK|$ is the size of a public key and $|Enc(K)|$ is the size of the encrypted symmetric encryption key.

\begin{figure}
\includegraphics[width=1.0\linewidth]{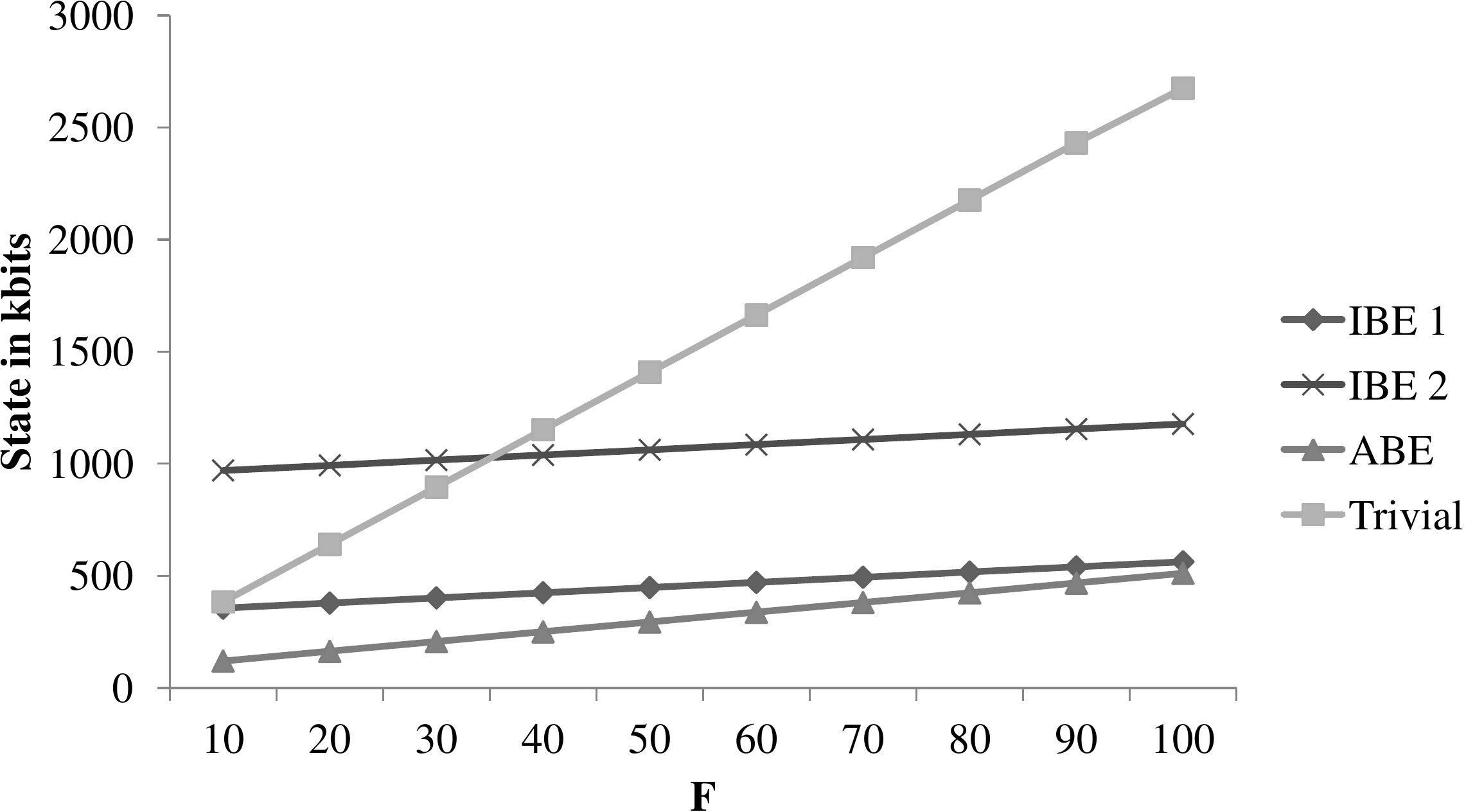}
\caption {Storage overhead as a function of $F$, the number of content items shared by each owner}
\label{fig:state2}
\end{figure}

When the ABE-based solution is used, each policy is treated as an attribute, 
therefore the storage node should maintain $U$ identities, $G \times U_G$ attribute identifiers, 
and $F \times U_G$ ABE encryptions of the symmetric encryption key. 
Therefore:
\begin{align}
\begin{split}
S = {} & U \times |ID| \\
& + G \times U_G \times |ID|\\
& +F \times (|ID|+|ABE_{attr}(K)|)
\end{split}
\end{align}
where $|ABE_{attr}(K)|$ is the size of the symmetric encryption key encrypted using ABE and a single attribute.
Figure~\ref{fig:state1} shows the storage overhead as a function of $U_G$. 
In this experiment $F$ is set to $50$.
Figure~\ref{fig:state2} shows the storage overhead as a function of $F$.
In this experiment $U_G$ is set to $25$.
In both experiments $U$ is set to $50$.

\subsubsection{Communication overhead}
We now examine the communication overhead
introduced when the right of a subscriber to access content is revoked. 
When our system is used,
the only thing that has to be done is to remove the identity of that subscriber from the
\emph{Access Control Policies} table.
Therefore, a content owner has to send a single message
to the storage node containing the changes to the access control policies. 
With the trivial solution, a content owner should remove the identity of
the revoked subscriber from the \emph{Access Control Policies} table
and
also
remove
all encryptions of the symmetric encryption key
generated using the public key of the revoked subscriber.
Finally, with the ABE-based solution, when a subscriber is revoked,
a content owner should remove the revoked subscriber from the
\emph{Access Control Policies} table,
re-generate the encryptions of the symmetric encryption key
(which
otherwise
the revoked subscriber could access),
and send the new ciphertexts to the storage node.

\section{Discussion}
\label{sec:discuss}
Our design allows each user to maintain his own PKG.
This feature protects the system from the key escrow problem and supports various other features based on  the properties of IBE.
For example, providing that a subscriber knows the $SP$ of a content owner, he can easily verify
that a storage node is allowed to store a content item (identifier),
since all cryptographic operations based on the item identifiers of the same owner use the same $SP$.
Another advantage of supporting per-user PKG is that key revocation becomes
easier: when per-user PKGs are used, users have the option to simply update their $SP$.
However, each user maintaining his own PKG creates an additional overhead for content owners
since they have to retrieve the $SP$ of their subscribers.
In section~\ref{setup} we mentioned solutions that can be used for secure $SP$ retrieval.
In any case, our scheme \emph{does not prohibit} the use of the same PKG among (a set of) users.
In that case PKGs know the users' secret keys,
but content owners only have to retrieve the $SP$ once for each set of users.
Moreover, providing that there is a secure method for a storage node to verify the identity of a subscriber,
the $SP$ of a subscriber (which are public) can be simply included in a message the subscriber sends to the node before any re-encryption.     

The authentication protocol presented in section~\ref{hand} allows subscribers to 
verify that a storage node is authorized to store a particular piece of content. 
This is achieved by encrypting a random number using the IBE \emph{encrypt} algorithm
with input the content item identifier. If a subscriber wants to receive multiple items
from a storage node and for each content item she wishes to verify that the node is authorized 
to store it, then this protocol has to be executed before every request. This may
result in increased network overhead and latency. An alternative approach 
could be to use the same prefix for a group of content item identifiers,
and verify that a storage node is allowed to store this prefix. In that case,
the subscriber should encrypt the random number using the prefix as input to
the IBE \emph{encrypt} algorithm.
Then, the authentication protocol has to be repeated only once,
when requesting items belonging to the same group.
Of course, this construction requires
content owners to generate the secret keys that correspond to identifier
prefixes and provide them in the storage nodes.

A feature not yet implemented in our system is content item listing.
Such listing function should respect the access control policies defined.
Note that in the scheme as described here,
access control policies are stored in the storage node and this may raise some privacy concerns;
extensions to our design that use other trusted parties for storing and evaluating access 
control policies (similar to~\cite{FMPCCR12}) can be introduced to address these concerns.
In that case the storage node would redirect subscribers requesting access to a content item
to that trusted (third) party,
which in turn would authenticate subscribers
and provide the storage node with the appropriate re-encryption key.

Finally, our design has been independent of any particular ICN architecture and applicable to all.
Only further optimizations, such as the prefix grouping discussed just above, depend on specific characteristics of the ICN architecture.
The key aspects of ICN that are exploited by our design are
the naming of individual information items by the content owner
and their atomic treatment,
at least at the level of users and storage nodes.

\section{Conclusion}
\label{sec:conc}
In this paper we presented a secure {ICN}-based content sharing system that leverages
identity-based proxy re-encryption.
Our scheme does not suffer from the key escrow problem,
it does not require any pre-shared secret information,
and it has low storage and network overhead.
Moreover, we
(1) presented an access control framework,
(2) designed an IBE-based authentication protocol limiting storage node and network traffic overhead by unauthorized users, and
(3) a construction that allows the use of untrusted storage nodes.

The security evaluation demonstrated that the scheme possesses the required properties
and in particular that storage nodes do not learn anything about subscriber secret keys or the shared content.
Even in the case of a storage node compromise, the outcome is reasonable and in particular
no information about the secret keys of the users is obtained.
The performance evaluation of the system illustrated the low storage and network overhead of the system and compared it to basic alternatives.

\section{Acknowledgments}
This research was supported by the EU funded H2020 ICT project POINT, under contract 643990. 

\bibliographystyle{acm}
\bibliography{icn-2016v6}

\end{document}